\documentclass[%
 aip,
 amsmath,amssymb,
 reprint,%
]{revtex4-1}

\usepackage{graphicx}
\usepackage{dcolumn}
\usepackage{bm}

\usepackage[utf8]{inputenc}
\usepackage[T1]{fontenc}
\usepackage{mathptmx}
\usepackage{etoolbox}
\usepackage{braket}
\usepackage[colorlinks=true,citecolor=blue]{hyperref}
\hypersetup{colorlinks=true,citecolor=blue,linkcolor=blue,urlcolor=blue}

\renewcommand{\i}{\ensuremath{\mathrm{i}}}
\newcommand{\e}{\ensuremath{\mathrm{e}}}

\makeatletter
\def\@email#1#2{%
 \endgroup
 \patchcmd{\titleblock@produce}
  {\frontmatter@RRAPformat}
  {\frontmatter@RRAPformat{\produce@RRAP{*#1\href{mailto:#2}{#2}}}\frontmatter@RRAPformat}
  {}{}
}%
\makeatother

\begin{document}

\title{{Phonon-assisted Auger decay of excitons in doped transition metal dichalcogenide monolayers}}

\author{Benedikt Scharf}
\affiliation{Institute for Theoretical Physics and Astrophysics and W\"{u}rzburg-Dresden Cluster of Excellence ct.qmats, University of W\"{u}rzburg, Am Hubland, 97074 W\"{u}rzburg, Germany}

\author{Vasili~Perebeinos}
\email{vasilipe@buffalo.edu}
\affiliation{Department of Electrical Engineering, University at Buffalo, Buffalo, NY 14228, USA}

\date{\today}

\begin{abstract}
The competition between the radiative and nonradiative lifetimes determines the optical quantum yield and plays a crucial role in the potential optoelectronic applications of transition metal dichalcogenides (TMDC). Here, we show that, in the presence of free carriers, an additional nonradiative decay channel opens for excitons in TMDC monolayers. Although the usual Auger decay channel is suppressed at low doping levels by the simultaneous momentum and energy conservation laws, exciton-phonon coupling relaxes this suppression. By solving a Bethe-Salpeter Equation, we calculate the phonon-assisted Auger decay rates in four typical TMDCs as a function of doping,  temperature, and dielectric environment.  We find that even at a relatively low doping of 10$^{12}$ cm$^{-2}$, the nonradiative lifetime ranges from 16-165 ps in different TMDCs, offering competition to the radiative decay channel.   
\end{abstract}

\maketitle
\section{Introduction}
The exceptional properties exhibited by two-dimensional (2D) materials have revolutionized the landscape of materials science and nanotechnology.\cite{Novoselov2016:S,Manzeli2017:NRM} The interplay between radiative and nonradiative recombination processes in transition metal dichalcogenides (TMDCs) is fundamental to their optoelectronic properties.\cite{Wang2012:NN,Mak2016:NP,Pospischil2016:AS} These two-dimensional semiconductors exhibit strong light-matter interactions,\cite{Britnell2013:S} making them promising candidates for various photonic and optoelectronic applications.\cite{Huang2022:RoPiP} The optical quantum yield,\cite{Amani2015:S,Lee2021:NC} a critical parameter for such applications, is directly influenced by the competition between the radiative\cite{Palummo2015:NL,Liu2019:JPCC,Robert2016:PRB,Ayari2018:PRB} and nonradiative\cite{Wang2024:NS} exciton lifetimes.\cite{Moody2016:JOSAB,Kuechle2021:OMX}

In TMDC monolayers, excitons, or bound electron-hole pairs, play a crucial role in determining the material's optical properties.\cite{Mak2010:PRL,Lin2014:NL,Wang2018:RMP,Regan2022:NMR} The radiative recombination of these excitons results in photon emission, which contributes to the material's luminescence. However, nonradiative recombination pathways can significantly impact the overall quantum efficiency of these materials.\cite{Peterson2024:arxiv} Therefore, understanding these dynamics is crucial for optimizing TMDC-based devices for various optoelectronic applications.\cite{Wang2012:NN,Mak2016:NP}

In this work, we demonstrate that the nonradiative exciton phonon-assisted Auger decay channel can effectively compete with the radiative decay even at moderately low doping levels, especially in smaller bandgap TMDCs. The conventional Auger decay mechanism, a nonradiative process often observed in semiconductors,\cite{Beattie1959:PRSL,Gelmont1978:JETP,Haug1983:JPC} is suppressed in TMDC monolayers at low doping concentrations. This suppression is attributed to the simultaneous conservation of momentum and energy in these 2D systems. However, phonon emission or absorption can relax momentum conservation, enabling phonon-assisted Auger decay of excitons available even at low doping levels.\cite{Perebeinos2010:PRL} The Auger decay with phonon assistance has been considered in bulk semiconductors.\cite{Bushick2023:PRL} In this work, we extend the phonon-assisted Auger decay mechanism to strongly bound excitons in 2D van der Waals materials. Strong exciton-phonon coupling in TMDCs\cite{Li2021:NC,Antonius2022:PRB,Katzer2023:PRB} is responsible, for example, for the temperature-dependent width of photoluminescent spectra\cite{Cadiz2017:PRX,Selig2016:NC,Raja2018:NL,Moody2015:NC,Brem2020:NL,Funk2021:PRR,Aslan2022:2DMat,Katzer2023:PRL} and magneto-optical spectroscopy\cite{Li2020:NC} and provides new possibilities for ultrafast intralayer-to-interlayer exciton transitions and interlayer charge transfer.\cite{Chan2024:NL}

\section{Methods and Model}
We use standard notations throughout this paper, where $e>0$ stands for the elementary charge, $\varepsilon_0$ is the vacuum permittivity, and $k_{\text{B}}$ is the Boltzmann constant.

\subsection{Excitons}
To compute the phonon-assisted Auger decay rate for excitons in TMDCs we first obtain the single-particle spectrum from the model tight-binding Hamiltonian derived in Ref.~\onlinecite{Fang2015:PRB}. Within the tight-binding model, a given single-particle state $\ket{n\bm{k}}$ with wave vector $\bm{k}$ in band $n$ and energy $\epsilon_{n\bm{k}}$ can be written as the Bloch sum
\begin{equation}\label{eq:BlochSum}
\ket{n\bm{k}}=\frac{1}{\sqrt{N}}\sum\limits_{\nu,\bm{R}}\e^{\i\bm{k}\cdot\bm{R}}a_{n\bm{k}\nu}\ket{\nu,\bm{R}},
\end{equation}
where $\ket{\nu,\bm{R}}$ is the orbital $\nu$ centered at the Bravais lattice point $\bm{R}$ and $N$ is the total number of primitive unit cells considered. The coefficients $a_{n\bm{k}\nu}$ in Eq.~(\ref{eq:BlochSum}) are determined from the tight-binding Hamiltonian via
\begin{equation}\label{eq:TBmodel}
\sum_{\nu'}\mathcal{H}_{\nu\nu'}(\bm{k})a_{n\bm{k}\nu'}=\epsilon_{n\bm{k}}a_{n\bm{k}\nu},
\end{equation}
where the 11-band tight-binding Hamiltonian $\mathcal{H}_{\nu\nu'}$ and its corresponding hopping parameters are taken from Ref.~\onlinecite{Fang2015:PRB}. For simplicity, we use the 11-band model without spin-orbit coupling in the remainder of this work.

The tight-binding states determined from Eqs.~(\ref{eq:BlochSum}) and~(\ref{eq:TBmodel}) are then used to compute the exciton dispersion via a Bethe-Salpeter equation (BSE). For an exciton $S$ with momentum $\bm{q}$, this BSE can be conveniently written as\cite{Rohlfing2000:PRB,Scharf2016:PRB}
\begin{eqnarray}
[\Omega_{S}(\bm{q})-\epsilon_c(\bm{k}+\bm{q})+\epsilon_v(\bm{k})] \mathcal{A}^{S\bm{q}}_{vc\bm{k}}\hspace{2.5cm}\nonumber\\
=\sum\limits_{v'c'\bm{k}'}\left[\mathcal{K}^\mathrm{d}_{vc\bm{k},v'c'\bm{k}'}(\bm{q})+\mathcal{K}^\mathrm{x}_{vc\bm{k},v'c'\bm{k}'}(\bm{q})\right]\mathcal{A}^{S\bm{q}}_{v'c'\bm{k}'},\label{eq:BSE}
\end{eqnarray}
where $\Omega_{S}(\bm{q})$ denotes the exciton energy and $\mathcal{A}^{S\bm{q}}_{vc\bm{k}}$ denote the coefficients of the linear combination forming the excitonic state
\begin{equation}\label{eq:ExcitonState}
\ket{\Psi_{S\bm{q}}}=\sum_{vc\bm{k}}\mathcal{A}^{S\bm{q}}_{vc\bm{k}}\hat{c}^\dagger_{c\bm{k}+\bm{q}}\hat{c}_{v\bm{k}}\ket{\mathrm{GS}}\equiv\hat{X}^\dagger_{S\bm{q}}\ket{\mathrm{GS}}.
\end{equation}
Here $\hat{c}^\dagger_{c\bm{k}}$ ($\hat{c}_{v\bm{k}}$) is the creation (annihilation) operator of an electron with momentum $\bm{k}$ in a conduction band $c$ (valence band $v$) and $\ket{\mathrm{GS}}$ is the ground state with fully occupied valence bands and unoccupied conduction bands, while $\hat{X}^\dagger_{S\bm{q}}$ is the creation operator of an exciton in state $S$ with momentum $\bm{q}$. Equation~(\ref{eq:BSE}) contains the direct and exchange terms
\begin{eqnarray}\label{eq:DirectTerm}
\mathcal{K}^\mathrm{d}_{vc\bm{k},v'c'\bm{k}'}(\bm{q})=-\sum\limits_{\nu\nu'}a^*_{c\bm{k}+\bm{q}\nu}a_{c'\bm{k}'+\bm{q}\nu}a_{v\bm{k}\nu'}a^*_{v'\bm{k}'\nu'}\nonumber\\
\times\left[\frac{1}{N}\sum\limits_{\bm{R}}\e^{-\i(\bm{k}-\bm{k}')\cdot\bm{R}}\;W(|\bm{R}|)\right]
\end{eqnarray}
and
\begin{eqnarray}\label{eq:ExchangeTerm}
\mathcal{K}^\mathrm{x}_{vc\bm{k},v'c'\bm{k}'}(\bm{q})=\sum\limits_{\nu\nu'}a^*_{c\bm{k}+\bm{q}\nu}a_{v\bm{k}\nu}a_{c'\bm{k}'+\bm{q}\nu'}a^*_{v'\bm{k}'\nu'}\nonumber\\
\times\left[\frac{1}{N}\sum\limits_{\bm{R}}\e^{-\i\bm{q}\cdot\bm{R}}\;V(|\bm{R}|)\right],
\end{eqnarray}
respectively. Equations~(\ref{eq:DirectTerm}) and~(\ref{eq:ExchangeTerm}) have been derived assuming point-like orbitals. The direct and exchange terms are computed in the real space from the screened interaction and the bare Coulomb potentials $W(|\bm{R}|)$ and $V(|\bm{R}|)$, respectively. If we do not account for doping in the BSE, $W(|\bm{R}|)=V(|\bm{R}|)$. In 2D materials this potential can be well modeled by the Keldysh potential\cite{Keldysh1979:JETP,Cudazzo2011:PRB,Berkelbach2013:PRB}
\begin{equation}\label{eq:KeldyshPot}
W(|\bm{R}|)=\frac{e^2}{8\varepsilon_0 r_0}\left[H_0\left(\frac{\kappa|\bm{R}|}{r_0}\right)-Y_0\left(\frac{\kappa|\bm{R}|}{r_0}\right)\right],
\end{equation}
where $H_0$ is the Struve function, $Y_0$ the Bessel function of the second kind, $\varepsilon_0$ the vacuum permittivity, $\kappa$ the averaged relative permittivity of the dielectric background\footnote{The relative permittivity in our model is the average value computed from the relative permittivity above and below the TMDC monolayer, for example, $\kappa=(1+3.9)/2=2.45$ for a TMDC monolayer on top of a SiO$_2$ substrate (with air above the TMDC monolayer).} and $r_0$ the monolayer polarizability parameter.

\begin{figure}[htb]
\includegraphics[width=0.45\textwidth]{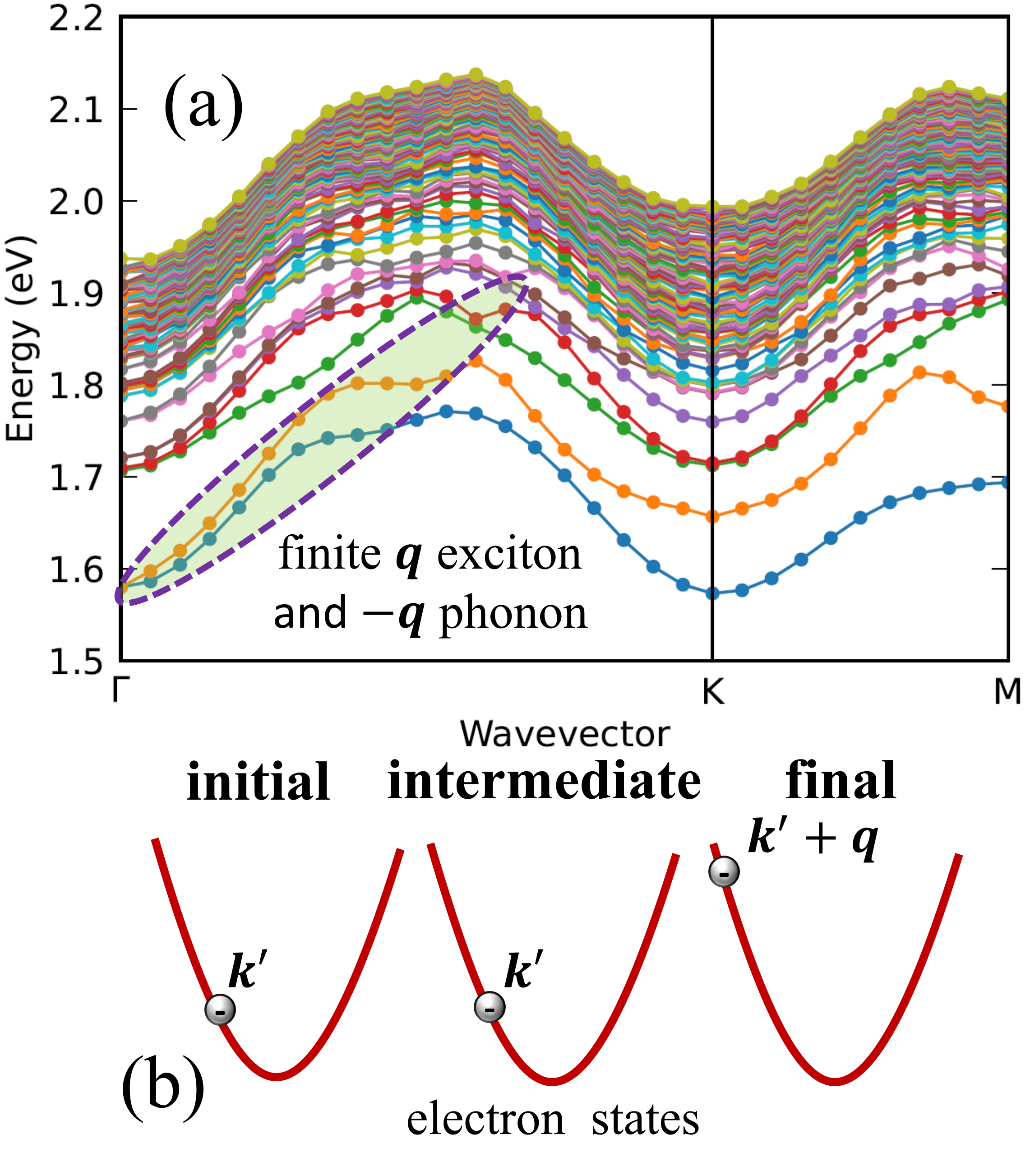}\\
\caption{Phonon-assisted Auger decay mechanism. (a) Exciton dispersion of MoSe$_2$ as a function of the center-of-mass momentum evaluated on a $60\times60$ $k$-mesh of the Brillouin  zone. Only the first 100 exciton bands are shown. The onset of the single-particle continuum at $q=0$ takes place at an energy of 1.89 eV. The colored oval demonstrates the transition of the zero-momentum exciton to a finite $q$-exciton by emission of a phonon with momentum $-\bm{q}$ in the intermediate state. (b) Electron states in the phonon-assisted Auger decay process: An initial state comprising a zero-momentum exciton and a conduction band electron with momentum $\bm{k}'$ is excited by emission (absorption) of a phonon with momentum $-\bm{q}$ ($\bm{q}$) into an intermediate state comprising an exciton with finite momentum $\bm{q}$ and the conduction band electron $\bm{k}'$. In the final state, the $\bm{q}$-exciton decays into the ground state by an Auger process and transfers its momentum to the conduction band electron.}\label{fig:Scheme}
\end{figure}

\subsection{Phonon-assisted Auger Decay Rate}
Having obtained $\Omega_{S}(\bm{q})$ and $\mathcal{A}^{S\bm{q}}_{vc\bm{k}}$ from Eq.~(\ref{eq:BSE}), we are now in a position to compute the phonon-assisted Auger scattering rate.

In the second-order processes we consider (see Fig.~\ref{fig:Scheme}), the initial state is given by a zero-momentum exciton and an electron with momentum $\bm{k}'$ in the lowest conduction band $c_0$. By emitting a phonon with momentum $-\bm{q}$ or absorbing a phonon with momentum $\bm{q}$, the exciton acquires finite momentum $\bm{q}$. This intermediate state, in turn, collapses to the final state by Auger decay of the exciton to the ground state. The energy conservation then requires $\Omega_S(\bm{0})+\epsilon_{c_0\bm{k}'}=\hbar\omega_{-\bm{q}}+\epsilon_{c'\bm{k}'+\bm{q}}$ for phonon emission and $\Omega_S(\bm{0})+\epsilon_{c_0\bm{k}'}+\hbar\omega_{\bm{q}}=\epsilon_{c'\bm{k}'+\bm{q}}$ for phonon absorption, where $\hbar\omega_{\pm\bm{q}}$ is the phonon dispersion and $\epsilon_{c\bm{k}}$ the electron dispersion in conduction band $c$. During the transition to the final state, we allow for excitations to an arbitrary conduction band $c'$, that is, not only within the lowest conduction band $c_0$ [Fig.~\ref{fig:Scheme}(b) shows only one conduction band for simplicity].

We model the interaction mediating the collapse from the initial state to the final state by the Hamiltonian $\hat{H}=\hat{H}_{\mathrm{e-ph}}+\hat{H}_{\mathrm{e-e}}$ with the exciton-phonon interaction
\begin{equation}\label{eq:ephInteraction}
\hat{H}_{\mathrm{e-ph}}=\frac{g}{\sqrt{N}}\sum\limits_{S\bm{k}\bm{q}}\hat{X}^\dagger_{S\bm{k}+\bm{q}}\hat{X}_{S\bm{k}}\left(\hat{b}^\dagger_{-\bm{q}}+\hat{b}_{\bm{q}}\right)
\end{equation}
and the electron-electron interaction
\begin{equation}\label{eq:eeInteraction}
\hat{H}_{\mathrm{e-e}}=\frac{1}{N}\sum\limits_{cc'v\bm{k}\bm{k}'\bm{q}}U^{c',c_0,c,v}_{\bm{k}'+\bm{q},\bm{k}',\bm{k}+\bm{q},\bm{k}}\hat{c}^\dagger_{c'\bm{k}'+\bm{q}}\hat{c}_{c_0\bm{k}'}\hat{c}_{c\bm{k}+\bm{q}}\hat{c}^\dagger_{v\bm{k}}.
\end{equation}
Equation~(\ref{eq:ephInteraction}) contains the exciton creation (annihilation) operator $\hat{X}^\dagger_{S\bm{k}}$ ($\hat{X}_{S\bm{k}}$) defined in Eq.~(\ref{eq:ExcitonState}). In our model the exciton-phonon coupling strength in Eq.~(\ref{eq:ephInteraction}) is given by a constant $g$, while the Auger Coulomb matrix element in Eq.~(\ref{eq:eeInteraction}) is given by
\begin{eqnarray}\label{eq:CoulombMatrixElement}
U^{c',c_0,c,v}_{\bm{k}'+\bm{q},\bm{k}',\bm{k}+\bm{q},\bm{k}}=\sum\limits_{\nu\nu'}a^*_{c'\bm{k}'+\bm{q}\nu}a_{c\bm{k}+\bm{q}\nu}a_{c_0\bm{k}'\nu'}a^*_{v\bm{k}\nu'}\nonumber\\
\times\left[\frac{1}{N}\sum\limits_{\bm{R}}\e^{-\i\left(\bm{k}'-\bm{k}\right)\cdot\bm{R}}\;V(|\bm{R}|)\right].
\end{eqnarray}
Here Eq.~(\ref{eq:CoulombMatrixElement}) has been calculated employing the same approximations as those used to arrive at Eqs.~(\ref{eq:DirectTerm}) and~(\ref{eq:ExchangeTerm}).

\begin{table}
\begin{center}
\begin{tabular}{|c||c|c|c|c|c|}
\hline
Material & $m_c/m_0$ & $m_v/m_0$ & a (\AA) & g (eV) \\
\hline\hline
MoS$_2$ & 0.47 & 0.57 & 3.18 & 0.211 \\
\hline
MoSe$_2$ & 0.54 & 0.65 & 3.32 & 0.176\\
\hline
WS$_2$ & 0.31 & 0.41 & 3.18 & 0.168\\
\hline
WSe$_2$ & 0.34 & 0.44 & 3.32 & 0.216 \\
\hline
\end{tabular}
\end{center}
\caption{Effective masses of the lowest conduction band and the uppermost valence band in units of the free electron mass $m_0$ (taken from Ref.~\onlinecite{Fang2015:PRB}). The primitive unit cell area is given by $A=a^2\sqrt{3}/2$, where $a$ is the lattice constant. The best-fit values of the exciton-phonon coupling g to Eq.~(\ref{eq:fit}) are also given in the last column.}\label{tab:effectiveMasses}
\end{table}

Using second order perturbation theory and Fermi's golden rule, we can determine the Auger scattering rate as
\begin{eqnarray}\label{eq:AugerDecay}
\frac{\hbar}{\tau}=\frac{2\pi g^2}{N^3}&&\sum\limits_{c',\bm{k}',S,\bm{q}}\frac{\left|\sum\limits_{cv\bm{k}}\mathcal{A}^{S\bm{q}}_{vc\bm{k}}U^{c',c_0,c,v}_{\bm{k}'+\bm{q},\bm{k}',\bm{k}+\bm{q},\bm{k}}\right|^2}{\left[\Omega_{S}(\bm{q})-\Omega_{S}(\bm{0})+\hbar\omega_{-\bm{q}}\right]^2}\nonumber \\
&&\times f(\epsilon_{c_0\bm{k}'})\left[1-f(\epsilon_{c'\bm{k}'+\bm{q}})\right]\left[1+n_{-q}\right]\nonumber\\
&&\times\delta\left[\epsilon_{c'\bm{k}'+\bm{q}}-\epsilon_{c_0\bm{k}'}+\hbar\omega_{-\bm{q}}-\Omega_{S}(\bm{0})\right]\nonumber \\
+\frac{2\pi g^2}{N^3}&&\sum\limits_{c',\bm{k}',S,\bm{q}}
\frac{\left|\sum\limits_{cv\bm{k}}\mathcal{A}^{S\bm{q}}_{vc\bm{k}}U^{c',c_0,c,v}_{\bm{k}'+\bm{q},\bm{k}',\bm{k}+\bm{q},\bm{k}}\right|^2}{\left[\Omega_{S}(\bm{q})-\Omega_{S}(\bm{0})-\hbar\omega_{\bm{q}}\right]^2}\nonumber \\
&&\times f(\epsilon_{c_0\bm{k}'})\left[1-f(\epsilon_{c'\bm{k}'+\bm{q}})\right]n_{q}\nonumber\\
&&\times\delta\left[\epsilon_{c'\bm{k}'+\bm{q}}-\epsilon_{c_0\bm{k}'}-\hbar\omega_{\bm{q}}-\Omega_{S}(\bm{0})\right],
\end{eqnarray}
where the first and second terms are the contributions due to phonon emission and absorption, respectively. 
The integral over $1/x^2$ in Eq.~(\ref{eq:AugerDecay}) should be evaluated as a principal part, that is, $1/x^2\to[1-\exp(-x^2/\sigma^2)]/x^2$, where the value of $\sigma$ depends on the $k$-mesh. Here we use $\sigma=50$ meV and a $60\times60$ $k$-mesh.
The functions
\begin{equation}\label{eq:Distr}
f(\epsilon)=\frac{1}{\exp\left[(\epsilon-\mu)/k_{\text{B}}T\right]+1};\; n_{\bm{q}}=\frac{1}{\exp(\hbar\omega_{\bm{q}}/k_{\text{B}}T)-1}
\end{equation}
denote the Fermi-Dirac and Bose-Einstein distribution functions with temperature $T$ and chemical potential $\mu$. In order to obtain Eq.~(\ref{eq:AugerDecay}), we have assumed that only the lowest conduction band $c_0$ s occupied by excess electrons\footnote{Strictly speaking, occupied conduction bands also lead to additional screening of $W(|\bm{R}|)$, resulting in $W(|\bm{R}|)\neq V(|\bm{R}|)$. For simplicity, we neglect this screening in $W(|\bm{R}|)$. Additional screening due to doping leads to smaller exciton binding energies as well as to a band gap renormalization, which mostly compensates for the reduction in binding energy. Since we only consider relatively small doping densities, we assume that these effects will not significantly affect the exciton wave function $\mathcal{A}^{S\bm{q}}_{vc\bm{k}}$.} and that phonon-assisted transitions can occur only between this band and any other conduction band.

\begin{figure*}
    \centering
\includegraphics[width=1.0\textwidth]{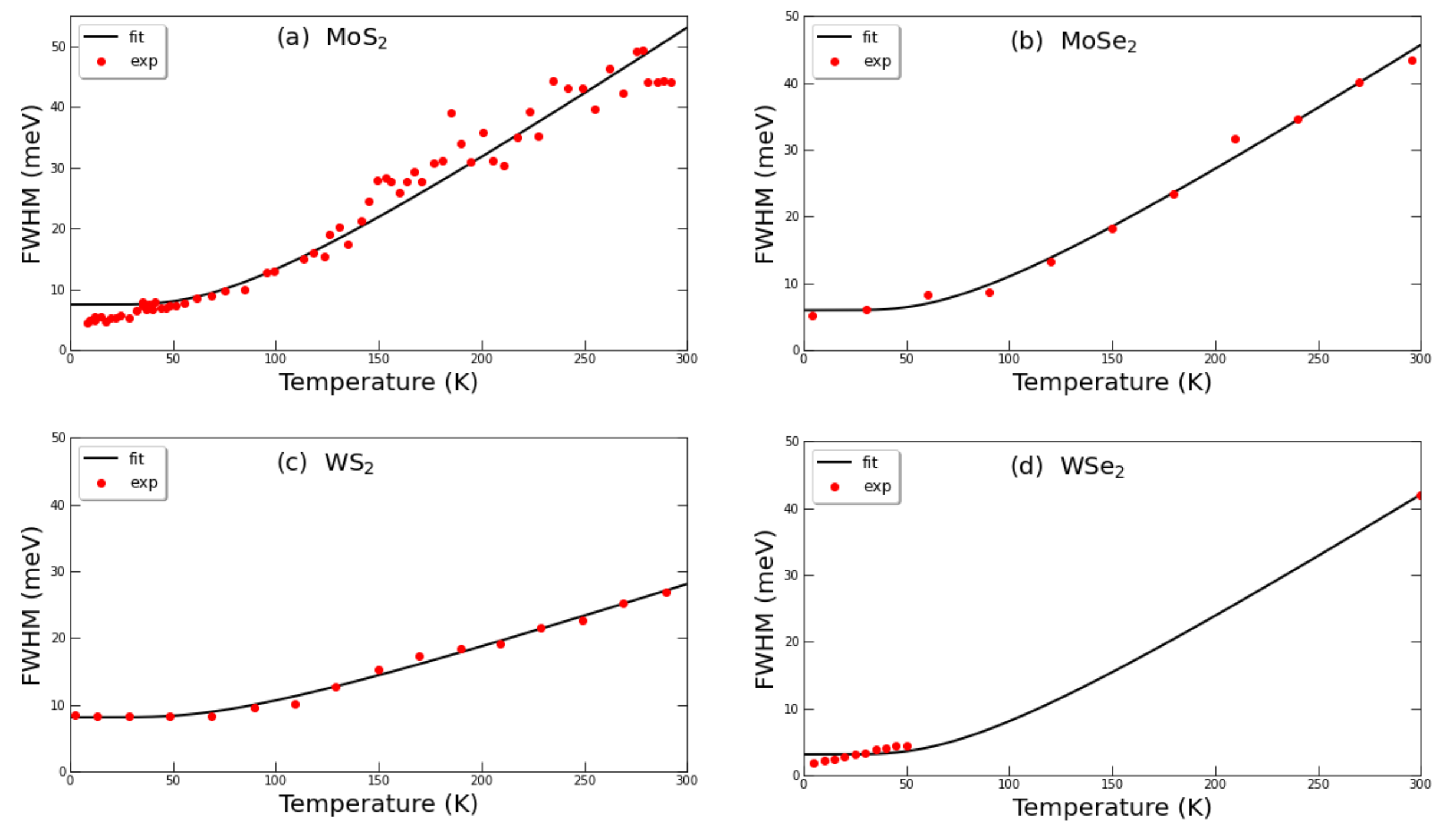}
\caption{Full-width at half maximum data analysis of (a) MoS$_2$,\cite{Cadiz2017:PRX} (b) MoSe$_2$,\cite{Selig2016:NC} (c) WS$_2$\cite{Raja2018:NL} and (d) WSe$_2$\cite{Moody2015:NC,Aslan2022:2DMat} to extract the corresponding strengths of the exciton-phonon coupling according to fits to Eq.~(\ref{eq:fit}).}\label{fig:fitting}
\end{figure*}

\subsection{Exciton-Phonon Coupling Strength}
To estimate the strength of the exciton–phonon coupling, we use the experimentally measured full-width at half-maximum $\gamma$ from temperature-dependent photoluminescence spectra (shown in Fig.~\ref{fig:fitting}) according to Fermi's golden rule,
\begin{eqnarray}\label{eq:fit}
\gamma(T)=\gamma_0+2\pi g^2 DOS_{exc} n_{ph}, 
\end{eqnarray}
where $\gamma_0$ is a temperature-independent broadening due to defects and sample inhomogeneities and the temperature dependence arises from the phonon population $n_{ph}$ according to Eq.~(\ref{eq:Distr}). The exciton density of states is evaluated according to $DOS_{exc}=m_{exc}A/(2\pi \hbar^2)$, where $A$ is the area of the primitive unit cell and the exciton mass $m_{exc}=m_c+m_v$ is given by the sum of conduction and valence band effective masses from Table~\ref{tab:effectiveMasses}. The values of $g$ from the best fits are given in Table~\ref{tab:effectiveMasses}. Note that we only consider phonon absorption in Eq.~(\ref{eq:fit}), since there are no states to scatter to as a result of the phonon emission for the lowest-energy excitons and we do not differentiate between acoustic and optical phonon contributions to the full-width at half maximum by using a fixed value of the phonon energy of $\hbar\omega_{ph}=20$ meV for all TMDCs considered in Eqs.~(\ref{eq:AugerDecay}) and~(\ref{eq:fit}). Note that we use a momentum- and band-independent approximation for the strength of the exciton-phonon coupling, which is in the spirit of the commonly used deformation potential approximation in the context of electrical transport. The deformation potential approximation is justifiable because of the weak dependence of the coupling between electrons and optical non-polar phonons. We also employ the Einstein model for the optical phonons, that is, a momentum-independent phonon energy. Since the parameters of our model are obtained by refitting the existing experimental data on phonon-induced broadening of photoluminescence spectra, we expect those are quantitative results, which are comparable with future experiments.

\begin{figure}[htb]
\includegraphics[width=0.5\textwidth]{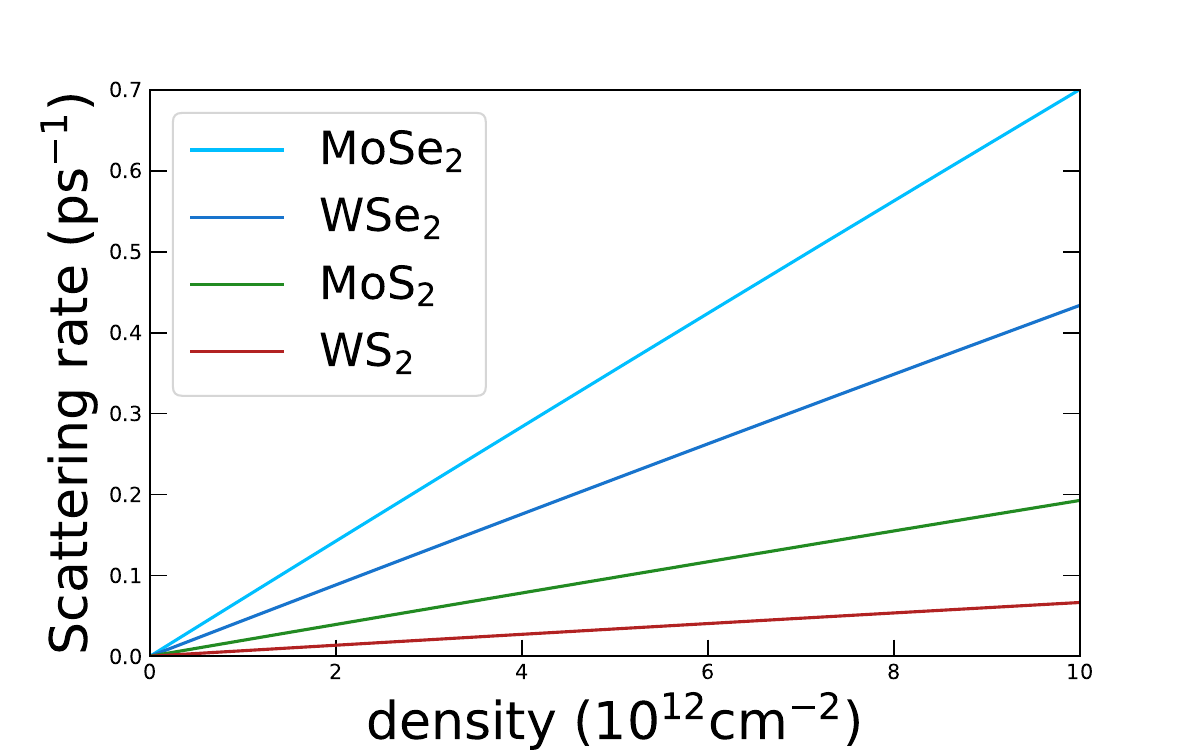}
\caption{ Phonon-assisted Auger decay rate of excitons in four most common TMDC monolayers as a function of doping, evaluated at $T=300$ K and in a dielectric environment $\kappa=3$.}\label{fig:Doping}
\end{figure}

\section{Results}
Equations~(\ref{eq:TBmodel})-(\ref{eq:fit}) can then be used to compute the phonon-assisted Auger decay rate. Here, we present the results for four different TMDCs, namely MoS$_2$, MoSe$_2$, WS$_2$ and WSe$_2$. As shown in Fig.~\ref{fig:Doping}, for all four TMDCs the scattering rates for the process considered in this work increase with doping densities and correspond to lifetimes in the sub-nanosecond regime at densities up to $10^{13}$ cm$^{-2}$.

\begin{figure}[htb]
\includegraphics[width=0.5\textwidth]{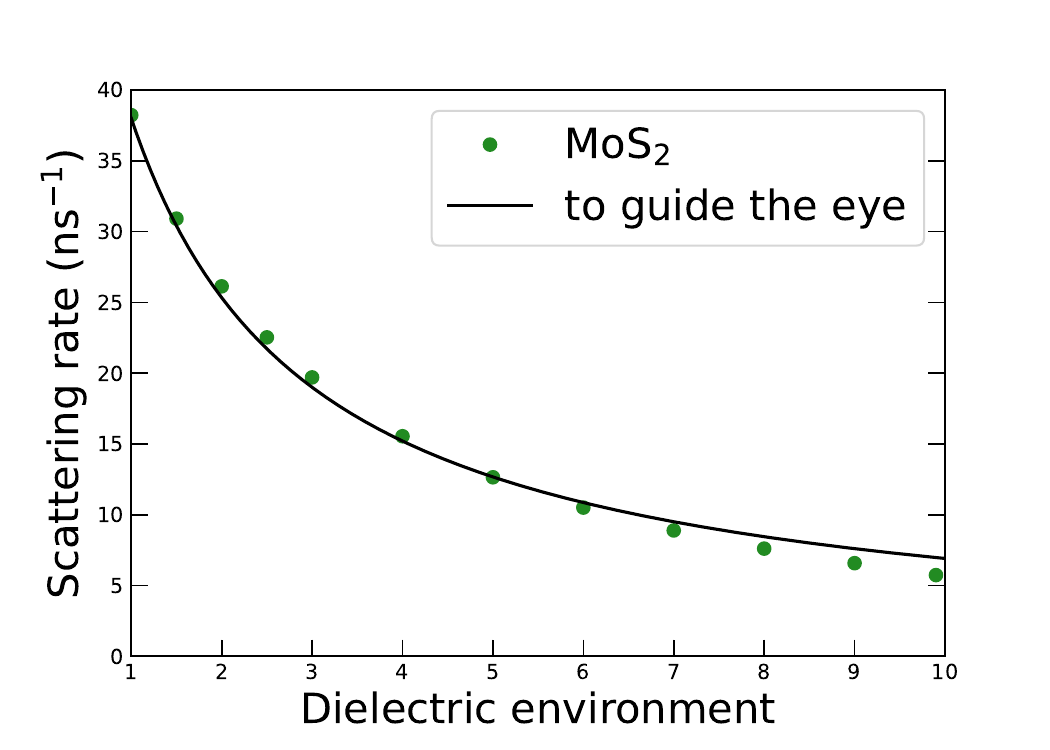}
\caption{Phonon-assisted Auger decay rate of excitons in MoS$_2$ as a function dielectric constant of the environment, evaluated at the doping level of $10^{12}$ cm$^{-2}$ and at $T=300$ K. The "to guide the eye" curve, $\propto1/(1+\kappa)$, demonstrates vanishing scattering rate in the limit of $\kappa\rightarrow\infty$.}\label{fig:Screening}
\end{figure}

In systems with strong dielectric screening, on the other hand, the phonon-assisted Auger scattering rate is suppressed, as illustrated in Fig.~\ref{fig:Screening}: The stronger the dielectric screening, the smaller the Auger scattering rate. In the limit of an infinitely large dielectric constant, the phonon-assisted exciton Auger decay rate vanishes as shown in Fig.~\ref{fig:Screening} containing data computed from Eq.~(\ref{eq:AugerDecay}) along with an empirical fit. 

The temperature dependence follows from the probabilities of phonon emission and absorption, that is, $1+2n_{ph}$. The overall rate depends sensitively on the bandgap of the TMDCs, that is, the smaller the bandgap, the stronger the phonon-assisted exciton Auger decay rate. To probe the proposed phonon-assisted exciton Auger decay mechanism, we want the experimental sample to be free of defects to limit other nonradiative defect-assisted decay channels. The gate can supply the free carrier in a TMDC in a controllable way, such that the measured phonon-assisted exciton Auger decay rate could be analyzed according to the following empirical formula: 
\begin{eqnarray}\label{eq:expfit}
\frac{1}{\tau}=\frac{1+2n_{ph}}{\tau_0}\frac{\rho}{\rho_0}\frac{\kappa_0+1}{\kappa+1},
\end{eqnarray}
where $\tau_0$ is the phonon-assisted exciton Auger decay lifetime measured at typical experimentally relevant conditions, that is, $\rho_0=10^{12}$ cm$^{-2}$, $\kappa_0=3$, and $T=4$ K. The value of $\tau_0$ falls in the range of 16 ps to 165 ps depending on the bandgap of the TMDC material. 

\section{Conclusions}
We have demonstrated that the phonon-assisted exciton Auger decay mechanism can effectively compete with the radiative decay channel in defect-free samples, even at moderately low doping levels. The predicted values of the phonon-assisted Auger lifetime for excitons under typical experimental conditions are predicted to be in the range of 16-165 ps depending on the bandgap of the TMDC material. We believe that our work will stimulate further experimental work.  

\begin{acknowledgments}
We gratefully acknowledge the support by the National Science Foundation under Grant No. 2230727 and the computational facilities at the Center for Computational Research at the University at Buffalo (\url{http://hdl.handle.net/10477/79221}).
\end{acknowledgments}

\bibliography{References}

\end{document}